\documentclass[aps,prl,twocolumn]{revtex4}
\usepackage{graphics}
\usepackage{graphicx}
\begin{document}
\title{Can vacancies lubricate dislocation motion in aluminum?}  
\author{Gang Lu and Efthimios Kaxiras}
\affiliation
{Department of Physics and Division of Engineering and Applied Science,\\
 Harvard University, Cambridge, MA 02138}
\begin{abstract} 
The interaction of vacancy with dislocations in Al
is studied using the Semidiscrete Variational Peierls-Nabarro 
model with {\it ab initio} 
determined $\gamma$-surface. For the first
time, we confirm theoretically the so-called vacancy 
lubrication effect on dislocation motion in Al, a discovery that 
can settle a long-standing controversy in dislocation theory for 
fcc metals. We provide insights on the lubrication effect 
by exploring the connection between dislocation mobility and its
core width. We predict an increased dislocation splitting in
the presence of vacancy. We find that on average there is a 
weak repulsion between vacancies and dislocations which 
is independent of dislocation character. 
\end{abstract}
\maketitle
Defects and their mutual interactions dominate the properties of 
materials that host them. Vacancies as point defects, 
have long been known to strongly interact with dislocations (line defects), 
and the study of
their interactions represents one of most challenging problems
in material science and engineering 
\cite{balluffi}. More than a decade ago, Benoit {\it et al.} 
discovered an interesting phenomenon in ultra-high-purity 
aluminum deformed at low temperature of 4.2 K.
They observed a marked decrease of elastic modulus in cold-worked Al, 
which they attributed to vacancy
enhanced dislocation mobility in Al, a novel phenomenon 
they termed dislocation lubrication effect \cite{lauzier,benoit}.
Corroborative experimental evidence led  
these authors to conclude that vacancies, 
generated from cold-work or irradiation, are solely responsible for 
the enhanced dislocation mobility. This indeed is a quite intriguing
result because traditionally vacancies are thought to lock dislocation
motion by forming atmospheres around the dislocation \cite{nabarro}. 
Furthermore the lubrication effect may hold the key to resolve
the long-standing controversy for Peierls 
stress ($\sigma_p$) estimated from plastic deformation and from internal
friction measurements. It is generally believed that kink pair formation
(KPF) of dislocations is responsible for the Bordoni peaks observed in 
internal friction measurements of fcc metals \cite{seeger,fantozzi}. 
However $\sigma_p$ derived from KPF mechanism 
is in the order of 10$^{-3} \mu$ ($\mu$ is the shear modulus) 
far greater than the critical resolved shear stress (CRSS) estimated 
from plastic deformation 
experiments, which is around 10$^{-5}$ to 10$^{-4}\mu$ \cite{seeger2}. 
This has been regarded as a serious problem because it casts doubt
on the well accepted KPF theory for the Bordoni peak in fcc metals. 
The controversy is particularly troublesome in light of the good 
agreement found in bcc and ionic crystals regarding $\sigma_p$ measured 
from internal friction and low temperature CRSS 
experiments \cite{fantozzi}. Therefore the vacancy lubrication 
effect may settle the controversy by proposing that vacancies strongly 
interact with dislocations and as a consequence lower their $\sigma_p$ 
to the level
that is consistent with the low temperature CRSS experiments and
thus bridge the gap in $\sigma_p$
\cite{lauzier,benoit}. Finally if the lubrication mechanism 
turns out to be general, it may lead to an innovation in molding
technology for materials with high $\sigma_p$ by introducing
vacancies to the materials \cite{kosugi}. However interesting the
lubrication effect may seem to be, it has not been widely accepted,
which in our opinion, is due to poor 
understanding of the phenomenon. In fact, there is no 
complete theoretical 
work ever published to address this problem, to the best of our 
knowledge. Therefore it is the purpose of 
this paper to present the first {\it ab initio} study of
the problem. As we will show in the following, our calculations not only 
provide theoretical support for the lubrication effect,
they also reveal other important difference in  
dislocation properties that are associated with the presence of 
vacancies.

In this paper, we employed the recently developed Semidiscrete Variational
Peierls-Nabarro (SVPN) model \cite{bulatov,lu1} in conjunction with 
{\it ab initio} determined $\gamma$-surfaces \cite{gamma}. The SVPN model 
provides an ideal framework for multiscale simulations of dislocation 
properties, and it combines an atomistic ({\it ab initio}) treatment of 
the interactions
across the slip plane and an elastic treatment of the continua on either
side of the slip plane.
The model has been shown to be quite successful in predicting
dislocation properties by comparing its predictions against the direct atomistic
simulations results \cite{bulatov,lu1}. For example, by using the $\gamma$-surface
calculated from Embedded Atom Method (EAM), we obtained $\sigma_p$ for dislocations 
in Al that are in excellent agreement with that from
direct atomistic simulations employing the same EAM potential \cite{lu1}.
More remarkably, a good agreement is also achieved for dislocations 
in Si \cite{bulatov} where the classic Peierls-Nabarro model fails 
owing to the its insufficiency to deal with narrow dislocations, 
such as dislocations in Si. Since one has no {\it a priori} knowledge 
regarding the size of dislocations in the presence of vacancies, the SVPN
model seems to be particularly useful to explore the interaction of
vacancy with dislocations. Besides 
$\sigma_p$, the model can also provide reliable results for other dislocation
properties, such as partial
separation distance and core width \cite{lu1}, as they are compared to direct
atomistic simulations \cite{bulatov2,fang}. Thus the strength
of this approach, when combined with {\it ab initio} calculations for
$\gamma$-surface
is that it produces essentially an atomistic simulation for dislocation 
properties without suffering from uncertainties associated with empirical
potentials.

In the SVPN approach, the equilibrium structure of a dislocation 
is obtained by minimizing the dislocation energy functional \cite{bulatov,lu1}
\begin{equation}
U_{disl} = U_{elastic} + U_{misfit} + U_{stress} + Kb^2{\rm ln}L,
\end{equation}
where
\begin{equation}
U_{elastic} = \sum_{i,j}\frac{1}{2}\chi_{ij}[K_e(\rho_i^{(1)}\rho_j^{(1)} +
\rho_i^{(2)}\rho_j^{(2)}) + K_s\rho_i^{(3)}\rho_j^{(3)}],
\end{equation}
\begin{equation}
U_{misfit} =  \sum_i\Delta x \gamma(\vec{f_i}),
\end{equation}
\begin{equation}
U_{stress} = - \sum_{i,l}\frac{x_i^2-x_{i-1}^2}{2}\rho_i^{(l)}\tau^{(l)},
\end{equation}
with respect to the dislocation Burgers vector density $\rho_i$.
Here, $\rho_i^{(1)}$, $\rho_i^{(2)}$ and $\rho_i^{(3)}$
are the edge, vertical and
screw components of the general Burgers vector density
defined at the $i$th nodal point as 
$\rho_i = (f_i - f_{i-1})/(x_i - x_{i-1})$,
where $f_i$ and $x_i$ are the displacement vector
and the coordinate of the $i$th nodal point (atomic row).
$\gamma(\vec{f}_i)$ is the
$\gamma$-surface that is determined from {\it ab initio}
calculations. 
$\tau^{(l)}$ is the external stress component interacting
with the corresponding Burgers vector density $\rho^{(l)}$.  
$\chi_{ij}$ is the discretized elastic energy kernel,
and $K$, $K_e$ and $K_s$ are the
pre-logarithmic elastic energy factors \cite{bulatov,lu1}.
The quantity $L$ entering the last term
is the outer cutoff radius for the configuration-independent
part of the elastic energy \cite{hirth}.
We identify the dislocation {\it configuration-dependent} 
part of the elastic energy and
the misfit energy as core energy, i.e., $U_{core} = U_{elastic} + U_{misfit}$.
The response of a dislocation to an external stress is achieved by minimization 
of the energy functional at
the given value of the applied stress.
An instability is reached when an optimal solution for the Burgers vector density 
distribution no longer exists, which
is manifested numerically by the failure of the minimization procedure to convergence.
$\sigma_p$ is then identified as the critical value of the applied stress
giving rise to this instability.        

In order to examine how vacancies change dislocation core structure by 
modifying atomic bonding across the
slip plane, we carry out {\it ab inito}
calculations for the $\gamma$-surface of Al with vacancies at the slip plane. 
Specifically, we select a supercell containing six Al layers in [111] 
direction with four atoms per layer, and remove one Al atom from the
top layer (right below the designated slip plane) of the supercell to simulate a 
vacancy concentration  at 4 at.\%. We should emphasize that 4 at.\%
represents the vacancy concentration at the dislocation
core region that we are interested, therefore it is much greater than the average
vacancy concentration of the bulk material.
The {\it ab initio} calculations are based on the pseudopotential 
plane-wave method with local density approximation \cite{kohn} to 
the exchange-correlation functional 
\cite{perdew}. A kinetic energy cutoff of 12 Ry for the plane-wave basis is used
and a $k$-point mesh consisting of (8,8,4) divisions along the reciprocal lattice
vectors is sampled for the Brillouin zone integration. Atomic relaxation is 
performed before we initiate the sliding. During
the sliding process, atoms are allowed to move only along [111] direction
while the atoms at the innermost two layers are held fixed. Volume relaxation is
also performed for each sliding distance to minimize the tensile stress on
the supercell. 

The {\it ab initio} determined $\gamma$-surface for Al with and without 
vacancies is presented in Fig. 1.  
In order to highlight
the vacancy effect on $\gamma$-surface, we also summarize in 
Table I some important stacking fault energies for both Al and Al+V 
(Al with vacancies) systems. These special stacking faults correspond to the
various extremes along [12$\bar{1}$] and [101] directions of the 
$\gamma$-surface. 
It is clear that the vacancy lowers the intrinsic and unstable 
stacking fault energy along [12$\bar{1}$] direction 
while increases the run-on stacking fault energy and unstable stacking
fault energy along [101] direction. Therefore it is not immediately
clear how dislocation core structure can be changed by vacancies, and
a detailed analysis based on SVPN model is needed.  
Since the experiments \cite{lauzier} have concluded that the 
change in elastic constants due to vacancies is not responsible for the observed
lubrication effect, we will simply use the experimental elastic constants 
of pure Al in our calculations for Al+V system \cite{lu1}. 

Having determined all the necessary parameters entering the model, we can
study the interaction of vacancy and dislocations using 
the SVPN model. We select four, namely, screw (0$^\circ$), 
30$^\circ$, 60$^\circ$ and edge (90$^\circ$) dislocations, 
all with the same Burgers vector, but different 
orientations, in our calculations. $\sigma_p$ calculated 
from the SVPN model for the dislocations with and without 
vacancies is listed in Table II. 
$\sigma_p$ represents the intrinsic mobility of a straight dislocation, 
and it relates to the kink pair formation energy, 2$W_k$, by 
\begin{equation}
2W_k = \frac{(16\mu b^3 a^3 \sigma_p)^{1/2}}{\pi},
\end{equation}        
where $b$ is the Burgers vector and $a$ the distance between 
neighboring Peierls valleys \cite{seeger,fantozzi,benoit2}. 
The corresponding values for Al are $\mu$ = 28.8 GPa, 
$b = 2.85$ \AA~ and $a = 2.47 $\AA. 
By measuring $2W_k$ (the activation energy for the Bordoni peak)
in internal friction experiments, one can derive $\sigma_p$
according to Eq. (5) for the relevant dislocation. 
For example, using the experimentally measured value of 
$2W_k$ (0.21 eV) for a screw dislocation, one obtains 
 $\sigma_p$ =  
8$\times 10^{-3}\mu$ (224 MPa) for the screw dislocation 
in pure Al \cite{fantozzi,benoit2}. This value of $\sigma_p$ is in
excellent agreement with our model result, 8.82$\times 10^{-3}\mu$
 for the same dislocation
(Table II). Furthermore, the less definite measurement for the 
subsidiary peak (B1 peak) yields an activation energy ranging from 
0.12 to 0.16 eV,  which corresponds to  $\sigma_p$ in the
range of 2.8 to 4.6$\times 10^{-3}\mu$ (80 to 130 MPa) for the 
60$^\circ$ dislocation. This value
also agrees well with our result for the same
dislocation (3.40$\times 10^{-3}\mu$). The overall consistency between the 
theoretical and experimental values for $\sigma_p$ indicates
the reliability of our model and establishes 
the basis for further study of vacancy effect. 

When vacancies are introduced at the slip plane but are not adsorbed
by a dislocation line, we find that $\sigma_p$ for various
dislocation is lowered by more than one order of magnitude (except
for the edge dislocation), as shown in Table II. Therefore we
have confirmed the vacancy lubrication effect theoretically 
for the first time since its proposal. 
The fact that this lubrication effect is observed for 
various dislocations suggests a generic nature of the 
underlying mechanism. In order to shed light on this general 
mechanism,
we have calculated dislocation core width which is defined
as the atomic spacing over which the relative displacement of the
dislocation changes from 1/4$b$ to 3/4$b$ \cite{lu1}.
It is generally believed that $\sigma_p$ is exponentially 
lowered with the increase of dislocation half-width according 
to the Peierls-Nabarro model \cite{hirth,lu1}.
The calculated dislocation half core width ($\zeta$) is 
presented in Table II. It is found that in the
presence of vacancy, dislocation becomes 60\% to 90\% wider, 
which we believe is due to the reduced slope of the $\gamma$-surface
along [12$\bar{1}$] direction as vacancies are 
introduced. Since lattice restoring
force, represented by the slope of $\gamma$-surface, is 
weakened by the vacancies, the repulsive elastic force 
resulting from the continuous distribution of infinitesimal 
dislocations dominates, leading to a wider dislocation core
and therefore enhanced dislocation mobility.
Although the vacancy lubrication effect may be qualitatively 
understood from above argument by a careful inspection of 
the $\gamma$-surface, one has to resort to 
SVPN model to obtain reliable values of 
$\sigma_p$ in order to make a quantitative comparison.
As shown in Table II, vacancies can bring $\sigma_p$
down to the values derived from the plastic deformation 
experiments (10$^{-5}$ to 10$^{-4} \mu$), therefore bridge 
the gap for $\sigma_p$ between the internal friction 
measurement and the measurement of CRSS  
at low temperature.   
  
In order to gain more insights into the interaction of vacancy 
and dislocations, we have calculated dislocation Burgers vector
density for pure Al and Al+V, shown in Fig. 2. It is found
that dislocations tend to dissociate more into partials in
the presence of vacancy. This behavior is obviously associated
with the fact that the intrinsic stacking fault energy is reduced
by vacancy. The result cautions us to be more careful in 
interpreting TEM data for partial separation distances because
accidentally introduced vacancies could change the result
significantly. We have also calculated
binding energies of vacancies to dislocation cores, 
summarized in Table II. The binding energy is defined as difference between
dislocation core energy with and without the presence of vacancy. 
Overall we find that the binding energies
are not sensitive to dislocation character,
and more importantly they are all marginally positive. 
The positive binding energy indicates that dislocation is 
energetically less stable with vacancies nearby. 
Our result for positive value of binding energy qualitatively
agrees with the findings from atomistic simulations carried out 
for Cu \cite{balluffi}. The small value of binding energy 
we found in Al may relate to the fact that the dislocations are
dissociated into partials in Al. It is believed that the vacancy 
binding energy to dissociated 
dislocations is considerably lower than that to
the same dislocations in the non-dissociated condition 
\cite{balluffi}. 
Of course one has to bear in mind that the binding energy
we obtained represents an average value over all possible sites
for a vacancy at the core of the dislocation. 

To conclude, we have studied the interaction of vacancy with
dislocations in Al using the SVPN model with {\it ab initio}
determined $\gamma$-surface. We confirm the experimental
finding of vacancy lubrication effect in Al. 
We propose that vacancies can weaken the
lattice restoring force across the slip 
plane, which leads to a wider dislocation spreading,
and  thus higher dislocation mobility. We find that $\sigma_p$
of the dislocations is lowered by more than one order of 
magnitude in the presence of vacancy, which bridges the
gap between $\sigma_p$ observed from different experiments, 
resolving one of the long-standing problems in dislocation 
theory. 
This work represents the first theoretical effort to
challenge the traditional point of view that regards vacancy as 
a locking agent for dislocation motion. 
We predict that vacancies can increase partial separation distance
in Al, and finally we find there exists a weak repulsion between
dislocations and vacancy, which is independent of 
the dislocation character.
   
\begin{acknowledgments}
We acknowledge the support from Grant No. F49620-99-1-0272
through the U.S. Air Force Office for 
Scientific Research.
\end{acknowledgments}

\begin{figure}
\includegraphics[width=300pt]{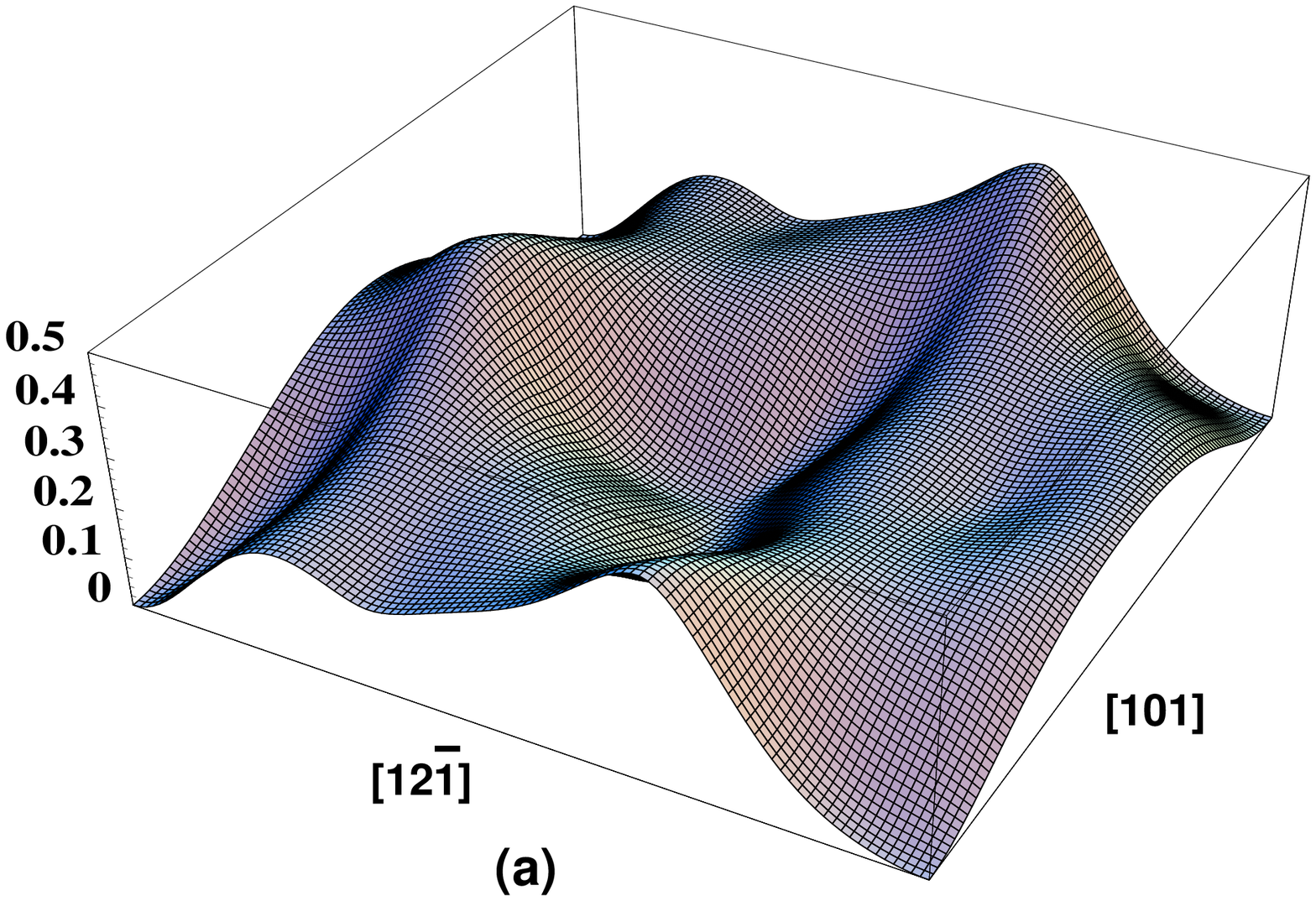}
\includegraphics[width=300pt]{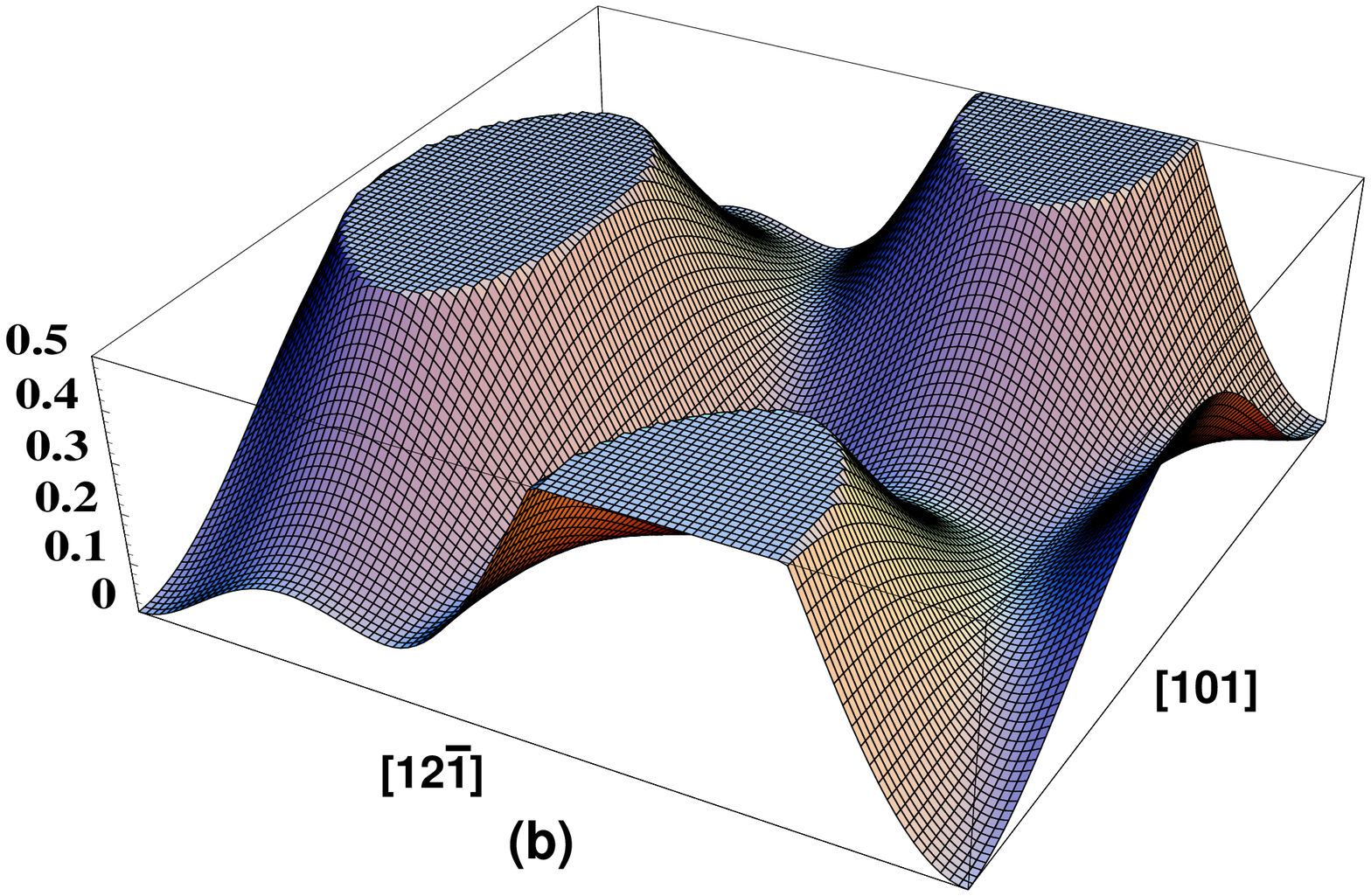}
\caption{The $\gamma$-surface (J/m$^2$) for displacements along a (111) plane 
for (a) pure Al and (b) Al+V systems.
The corners of the plane and its center correspond to identical equilibrium
configuration, i.e., the ideal lattice. The two surfaces are displayed in exactly
the same perspective and on the same energy scale to facilitate comparison.
The $\gamma$-surface of Al+V is truncated  
to emphasize the more interesting region.}
\end{figure}

\begin{figure}
\includegraphics[width=300pt]{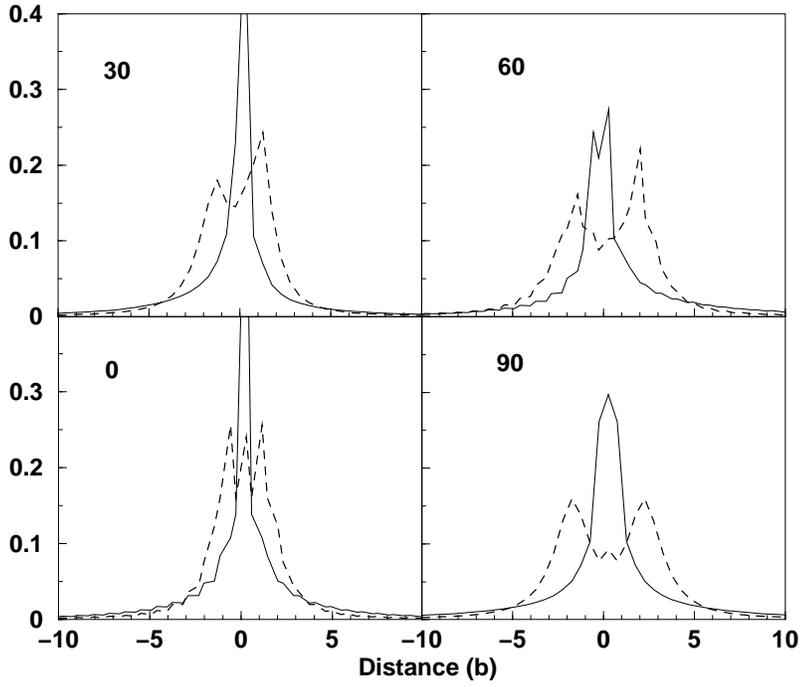}
\caption{Dislocation Burgers vector density for four dislocations (clockwise)
: screw (0$^\circ$), 30$^\circ$, 60$^\circ$ and edge (90$^\circ$) for the 
pure Al (solid lines) and the Al+V (dashed lines) systems. 
The peaks in the density plot represent partial dislocations.}
\end{figure}

\begin{table}[p]     
\caption{Fault vectors and energies (J/m$^2$) for some important stacking faults 
of the pure Al and the Al+V systems.}
\begin{ruledtabular}
\begin{tabular*}{\columnwidth}{@{\extracolsep{\fill}}cccc}
                    & Vector & Al  &  Al+H \\ \hline
 Intrinsic stacking & 1/6[12$\bar{1}$]  & 0.164 & 0.105 \\     
 Unstable stacking  & 1/10[12$\bar{1}$] & 0.224 & 0.143 \\
 Unstable stacking  & 1/4[101] & 0.250 & 0.427 \\
 Run-on stacking    & 1/3[12$\bar{1}$]  & 0.400 & 0.831\\       
\end{tabular*}     
\end{ruledtabular}        
\end{table}

\begin{table}[p]          
\caption{Peierls stress ($\sigma_p$, 10$^{-3}\mu$), half core width ($\zeta$,
 \AA),
core energies ($U_{core}$, eV/\AA)
for the four dislocations               
in the pure Al and the Al+V systems and
binding energy ($U_{b}$, eV/vacancy) for the four dislocations.}
\begin{ruledtabular}
\begin{tabular*}{\columnwidth}{@{\extracolsep{\fill}}cccccc}

  &       & screw &  30$^\circ$  & 60$^\circ$ & edge \\ \hline   
$\sigma_p$ & Al & 8.82  & 1.77 & 3.40 & 0.10 \\ 
           & Al+V & 0.69 & 0.10 & 0.26 & 0.05 \\ \hline
$\zeta$    & Al & 2.1   &2.5  & 3.0 & 3.5 \\
           & Al+V& 3.2   &4.0  & 5.6 & 6.3 \\ \hline
$U_{core}$ &  Al& -0.084 & -0.110 &  -0.168    & -0.198 \\
           & Al+V&-0.046 & -0.073 & -0.133 & -0.164 \\ \hline
$U_{b}$    &     &0.038 & 0.037  & 0.035 & 0.034 \\   
\end{tabular*}
\end{ruledtabular}
\end{table}   


\begin{thebibliography}{99}
\bibitem{gamma}
When a crystal is cut along its slip plane and the upper half is displaced
relative to the lower by a vector $\vec{f}$, the energy increase per unit area
is defined as $\gamma$ and the energy surface 
$\gamma(\vec{f})$, with $\vec{f}$ spanning the entire slip plane, is 
called the $\gamma$-surface.
\end{thebibliography}
\end{document}